\documentstyle[epsf,editedvolume,numreferences]{crckapb}

\newcommand{\beq}{\begin{equation}}
\newcommand{\eeq}{\end{equation}}
\newcommand{\bea}{\begin{eqnarray}}
\newcommand{\eea}{\end{eqnarray}}
\newcommand{\bi}{\begin{itemize}}
\newcommand{\ei}{\end{itemize}}
\newcommand{\be}{\begin{enumerate}}
\newcommand{\ee}{\end{enumerate}}
\newcommand{\bd}{\begin{description}}
\newcommand{\ed}{\end{description}}
\newcommand{\bfig}{\begin{figure}}
\newcommand{\efig}{\end{figure}}

\newcommand{\phrb}{Phys. Rev. B } 
 
\newcommand{\phrl}{Phys. Rev. Lett. }

\begin{opening}

\title{FINITE SIZE EFFECTS IN SMALL PARTICLE SYSTEMS}
\author{\`Oscar Iglesias, F\`elix Ritort, and Am\'{\i}lcar Labarta}
\institute{Departament de F\'{\i}sica Fonamental\protect\\
Universitat de Barcelona\protect\\ 
Diagonal 647, 08028 Barcelona, Spain}
\end{opening}

\begin{document}
\begin{abstract}
We present the results of Monte Carlo simulations of a model
of a $\gamma$-Fe$_2$ O$_3$ (maghemite) single particle of spherical shape. 
The magnetic Fe$^{3+}$ ions are represented by Ising spins on a spinel 
lattice that consists on two sublattices with octhaedral and tetrahedral 
coordination with exchange interactions among them and with an external
magnetic field. By varying the particle diameter, we have studied the 
influence of the finite size of the particle on the equilibrium properties, 
field cooling magnetization and hysteresis loops. The simulations allow to distinguish
the different roles played by the surface and the core spins of the particle
on its magnetic properties. We show that for small particle sizes the core 
is uncoupled from the surface, that behaves as a quasi-independent layer, whereas for
bigger particles the surface and the core are coupled and follow the behaviour of the 
bulk. 
\end{abstract}
%-------------------------------------------------------------------------------
%		Introduction
%-------------------------------------------------------------------------------

\section{INTRODUCTION}
Recent experimental studies in small particle systems of nanometric size 
have brought renewed interest in these kind of systems because of their 
anomalous
magnetic properties at low temperatures. Among the static properties, 
experiments have shown that in these systems the hysteresis loops display 
high closure fields and do not saturate even at fields of the order of 50 Tesla 
\cite{Kodama97,Garcia99,Martinez98} which indicates that the anisotropy fields cannot be the only 
responsible mechanism for the magnetization reversal. Low magnetization as 
compared to bulk, shifted loops after field cooling and irreversibilities 
between the field cooling and zero field cooling processes even at high fields 
are also observed \cite{Garcia99,Martinez98}. On the other hand, the time-dependence of the 
magnetization, in particular the existence of aging phenomena \cite{Jonsson95}, indicates 
that there must be some kind of freezing preventing the system to evolve towards 
equilibrium. 

Whether these phenomena can be ascribed to intrinsic properties 
of the particle itself (surface disorder which creates an exchange field on the 
core of  the particle), or they are due to a collective effect induced by  
interparticle interactions \cite{Batlle97,Dormann98,Morup94} has been the object 
of controversy in recent years 
and up to the moment there is no model giving a clear-cut explanation of this 
phenomenology although simulation results for general small particle 
systems \cite{Trohidou,Dimitrov94,Dimitrov95} and in particular for maghemite
\cite{Kach00,Kodama99} have been recently published . 
In order to elucidate this controversy we present the results of 
a Monte Carlo simulation of a single spherical particle which aims to clarify 
what is the specific role of the finite size and the surface on the magnetic 
properties of the particle, disregarding the interparticle interactions effects.

%-------------------------------------------------------------------------------
%		Model
%-------------------------------------------------------------------------------
\section{MODEL}
$\gamma$-Fe$_2$ O$_3$, maghemite, is one of the most commonly studied 
nanoparticle compounds \cite{Kodama99} presenting the above mentioned phenomenology. 
Maghemite is a ferrimagnetic spinel in which the magnetic Fe$^{3+}$ ions with spin 
$5/2$ are disposed in two sublattices with different coordination with the 
O$^{2-}$ ions. Each unit cell has 8 tetrahedric (T) and 16 octahedric (O) sites 
and in the real material one sixth of the O sites has randomly distributed 
vacancies. Thus, the spins in the T sublattice have $N_{TT}=4$ nearest 
neighbours in T and $N_{TO}=12$ in O and the spins in the O sublattice have 
$N_{OO}=6$ nearest neighbours in O and $N_{OT}=6$ in T.  
In our model the Fe$^{3+}$ magnetic ions are represented by Ising spins 
$S_i^{\alpha}$ distributed in two sublattices $\alpha= T,O$ of linear size $N$ 
unit cells, thus the total number of spins is ($24N^3$). The choice of Ising spins 
allows to reproduce a case with strong uniaxial anisotropy while keeping 
computational efforts within reasonable limits.
%----------------------------FIG.1---------------------------------------------
\bfig[tbp]
\centering
\epsfxsize=13.7 cm 
\leavevmode 
\epsfbox{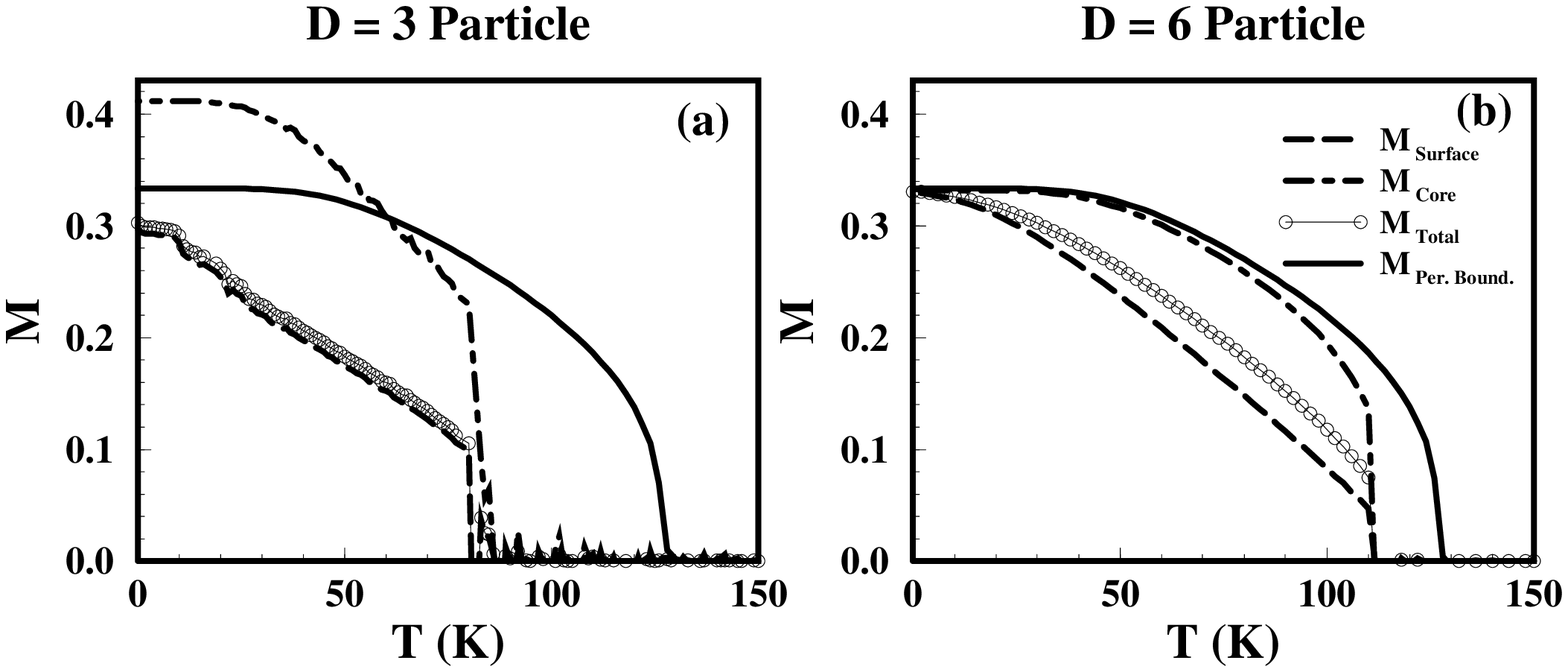} 
%\special{ psfile=Hysteresis.ps
%          angle=-90 hscale  = 40   vscale  = 40
%                    voffset = +220 hoffset = +25 }
\caption{Thermal dependence of the equilibrium magnetization for a spherical 
particle of (a) diameter D= 3 unit cells and (b) D= 6. The results corresponding to
a bulk system of the same size but periodic boundary conditions is also displayed
as a continuous line. Surface (dashed lines) and core (dot-dashed lines) contributions 
are distinguished from the total magnetization (open circles). $M$ has been normalized
to the number of surface, core or total spins depending on the case.
} 
\label{Fig1}
\efig
%----------------------------FIG.1---------------------------------------------

The spins interact via antiferromagnetic exchange interactions with the nearest 
neighbours on both lattices and with an external magnetic field $H$, and the 
corresponding Hamiltonian of the model is
\bea
%{\cal H}= -J_{TT}\sum_{i=1}^{N_T}\sum_{n=1}^{N_{TT}} S_i^{T} S_{i+n}^{T}
%          -J_{OO}\sum_{i=1}^{N_O}\sum_{n=1}^{N_{OO}} S_i^{O} S_{i+n}^{O}
%          -J_{TO}\sum_{i=1}^{N_T}\sum_{n=1}^{N_{TO}} S_i^{T} 
%S_{i+n}^{O}\nonumber\\
%          -J_{OT}\sum_{i=1}^{N_O}\sum_{n=1}^{N_{OT}} S_i^{O} S_{i+n}^{T}
%          -H\sum_{\alpha= T,O}\sum_{i=1}^{N_\alpha} S_i^{\alpha}
{\cal H}/k_{B}= -\sum_{\alpha,\beta=\,T,O}\sum_{i=1}^{N_\alpha}\sum_{n=1}^{N_{\alpha\beta}}
          J_{\alpha\beta} S_i^{\alpha} S_{i+n}^{\beta}
         -h\sum_{\alpha= T,O}\sum_{i=1}^{N_\alpha} S_i^{\alpha}\ .
\eea
The reduced magnetic field $h=\mu H/k_{B}$ is in temperature units.   
The values of the nearest neighbour exchange constants for maghemite are \cite{Kodama99} 
$J_{TT}=-21\  K$, $J_{OO}= -8.6\ K$, $J_{TO}= -28.1\ K$. We have used periodic 
boundary conditions to simulate the bulk properties and free boundaries for a 
spherically shaped particle with $D$ unit cells in diameter when studying finite 
size effects. In the latter case two different regions are distinguished in the 
particle: the surface formed by the outermost unit cells and an internal core 
of diameter $D_{Core}$ unit cells. The size of the studied particles ranges from 
$D= 3$ to $6$ corresponding to real particle diameters from 30 to 50 \AA. 

\section{RESULTS and DISCUSSION}
In Fig. 1 we present the thermal dependence of the total equilibrium 
magnetization for a system with periodic boundaries and two different particle 
sizes. First of all we note that the transition temperature does not depend 
significantly on the system size and has roughly the experimental value of the 
bulk (T$_c \simeq$ 900 K), taking into account that in the simulations the value 
of the spin has been set to 1 instead of 5/2. On the contrary, the ordering 
temperature for the spherical particle with free boundaries decreases with 
decreasing particle size. For the D=3 particle the total magnetization is dominated by the 
contribution of the surface, and the core, which includes only few spins, is 
almost uncoupled from the surface and follows the bulk behaviour. For D=6, the 
core and the surface are more coupled and the total magnetization is an 
average of surface and core contributions. In agreement with experimental 
observations, $M_{Total}$ increases with decreasing particle size at a finite T, tending 
to the bulk value.    
%----------------------------FIG.2---------------------------------------------
\bfig[tbp]
\centering
\epsfxsize=13.7 cm 
\leavevmode 
\epsfbox{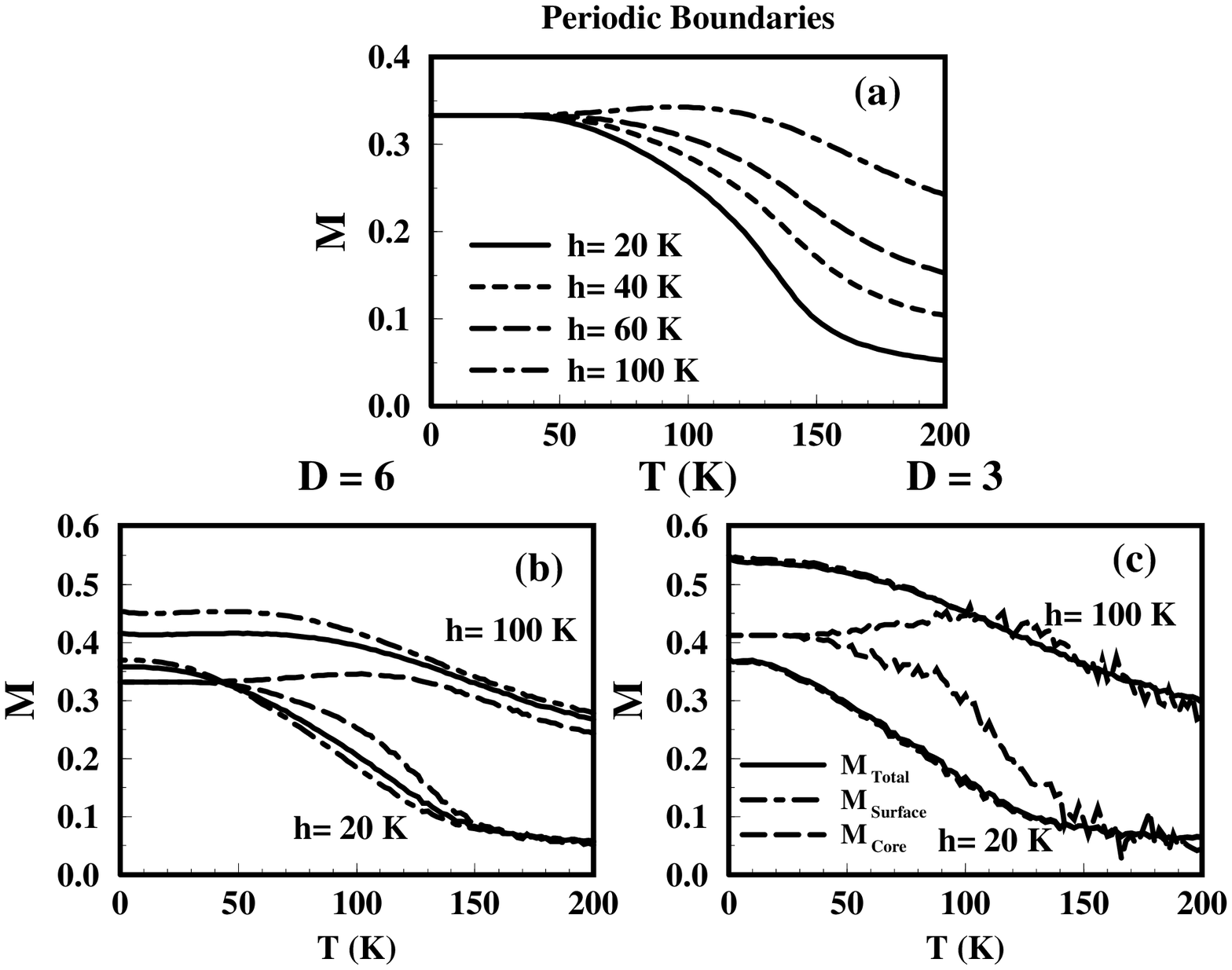} 
\caption{Thermal dependence of the magnetization after cooling under different fields 
displayed in the legends. (a) Shows the bulk results obtained for a system of linear 
size D= 8 and periodic boundaries. Figures(b) and (c) correspond to FC processes at
low $h= 20 K$ and high $h= 100 K$ fields for spherical particles
of diameters D= 3, 6 respectively. The contributions of the surface and the core 
have been distinguished as displayed in the legend. Reduced fields $h$ can be 
translated to $Oe$ multiplying by $k_{B}/\mu$.
} 
\label{Fig2}
\efig
%----------------------------FIG.2---------------------------------------------

To see what is the effect of a magnetic field on the magnetic order we have 
studied the field cooled (FC) magnetization at different cooling fields. The 
simulation procedure has been started from a disordered state at a temperature 
well above T$_c$ which is progressively reduced at a constant rate. The results 
for periodic boundaries show that
at high enough fields there is a maximum in the FC curve slightly below T$_c$ 
that is due to the competition between the ferromagnetic order induced by the 
field and the spontaneous ferrimagnetic order below T$_c$ (see Fig. 2a). This 
feature is also observed for the spherical particles but progressively 
disappears as the particle size is decreased (see Figs. 2b and 2c). This fact is 
much more pronounced in the core than in the surface due to the frustration 
induced by broken links at the particle boundary. As the particle size is 
decreased, the total magnetization becomes dominated by the contribution of the 
surface spins independently of the magnetic field (note that the continuous and 
dot-dashed lines in Fig. 2c superimpose for the D= 3 particle).
%----------------------------FIG.3---------------------------------------------
\bfig[htbp]
\centering
\epsfxsize=13.7 cm 
\leavevmode 
\epsfbox{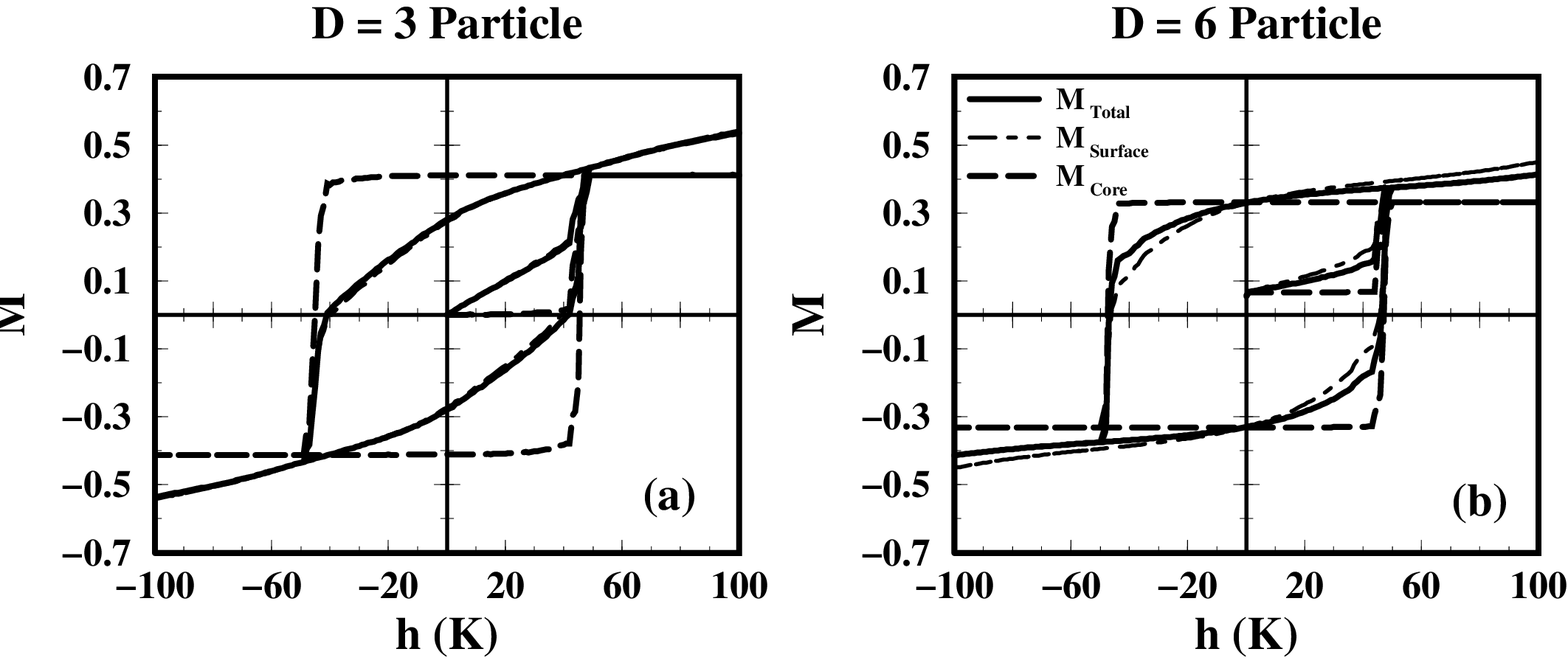} 
\caption{Hysteresis loops for a particles of diameter (a) D= 3 unit cells and (b)
D=6. The hysteresis loops of the surface (dot-dashed line) and core (dashed line)
have been plotted separatedly from the total magnetization (continuous line). 
} 
\label{Fig3}
\efig
%----------------------------FIG.3---------------------------------------------
Independently of the particle size and the cooling field, the core magnetization tends to the bulk 
value at low T because below T$_c$ the core tends to order ferrimagnetically. The shape of the 
FC curves for the core resemble the one for periodic boundaries even for the smallest 
particle, where the total magnetization is competely dominated by the surface and 
the core contains only 5\% of the total number of spins. However, this phenomenon is not observed
in the surface contribution due to the fact that the FC strongly depends on the 
magnetic field even at temperatures below T$_c$. 
The ferrimagnetic order is less perfect in the surface and the magnetic field 
induces ferromagnetic ordering at much lower field values.

Finally, we have simulated hysteresis loops for two particle sizes at a 
temperature $T= 20 K$ well below T$_c$ (see Fig. 3). First of all, let us note that 
the saturation field and the high field susceptibility increase as the particle 
size is reduced (compare the continuous lines of Fig. 2a and Fig. 2b) reducing the 
squareness of the loops, an effect 
that is due to the increasing importance of the surface contribution in the smallest 
particles. 
The hysteresis loop of the core is squared for both particle sizes: the core 
suddenly reverses as a whole, coherently rotating at the coercive field. In 
contrast, the surface hysteresis loop has a shape similar to the total 
magnetization loop, closing at the same field than the core but with coercive 
fields lower than the closure field as the particle size decreases.
This indicates that the surface gradually reverses its magnetization and that the 
surface spins rotation is not coherent as in the core. 

\paragraph{Acknowledgements} We acknowledge CESCA and CEPBA under coordination of 
C$^4$ for the computer facilities. This work has been supported by CICYT through 
project MAT97-0404 and CIRIT under project SR-119.

\end{document}